\documentclass[aps,amsmath,amssymb]{revtex4}

\usepackage{graphicx}
\usepackage{bm}

\newcommand{\be}{\begin{equation}}
\newcommand{\ee}{\end{equation}}
\newcommand{\bea}{\begin{eqnarray}}
\newcommand{\eea}{\end{eqnarray}}
\newcommand{\nn}{\nonumber\\}

\newcommand{\pa}[1]{\left(#1\right)}
\def\eq#1{(\ref{#1})}
\def\ord#1{{\cal O}(#1)}

\def\tr{{\mathrm Tr}}
\def\psid{\psi^\dagger}
\def\Psid{\Psi^\dagger}
\def\cd#1{{\cal D}[#1]}
\newcommand{\fd}[2]{\frac{\delta #1}{\delta #2 }}
\newcommand{\fdd}[3]{\frac{\delta^{2} #1}{\delta #2 \delta #3}}
\def\v#1{{\bm#1}}
\def\sp{\shortparallel}
\def\la{\langle}
\def\ra{\rangle}

\begin{document}

\title{Internal space renormalization group for the Luttinger model}
\author{Janos Polonyi$^{a,b}$}\email{polonyi@fresnel.u-strasbg.fr}
\author{Franck Stauffer$^a$}\email{stauffer@lpt1.u-strasbg.fr}
\affiliation{$^a$Theoretical Physics Laboratory, CNRS and
Louis Pasteur University, Strasbourg, France}
\homepage{http://lpt1.u-strasbg.fr}
\affiliation{$^b$Department of Atomic Physics,
L. E\"otv\"os University, Budapest, Hungary}

\date{\today}
\begin{abstract}
The absence of fermionic, asymptotical one-particle states in the Luttinger model raises
the suspicion that the interactions are actually strong at the vicinity of the
Fermi points. The functional internal space renormalization group method,
a systematical scheme for the computation of effective coupling
strengths off the Fermi points, is applied to shed some light on this issue.
A simple truncation of the evolution equation shows that the theory
is indeed strongly coupled at the Fermi points for arbitrarily small 
value of the bare coupling constant and the removal of the cutoff
is blocked by Landau poles. A peculiar feature of the normal ordering
scheme is pointed out to explain the absence of these effects in the
bosonized solution.
\end{abstract}
\maketitle

\section{Introduction}
One dimensional massless fermions display rather peculiar features.
Owing to another specialty of one dimension, to the equivalence of
the helicity and the chirality, the left and right moving
fermion numbers are individually conserved in chiral invariant
theories, like in the massless Thirring \cite{thirr} or Gross-Neveu
\cite{gn} models. The physics of electrons moving in a one-dimensional
wire can effectively be described by linearizing the dispersion relation
around the Fermi points. The resulting Luttinger model \cite{lut},
a theory of relativistic, massless fermions, is the
massless Thirring model except that the dispersion relations of the
left and right moving fermions are displaced horizontally with
respect to each other in order to accommodate the chemical potential.
This model provides a rich structure to compare with experimentally
well controlled situations \cite{varma}. 

One dimensional models can elegantly be renormalized by means of
normal ordering and the works aiming at the Luttinger model 
are almost exclusively based on this method. Most widespread
use of normal ordering is the bosonization. The bosonization of the
excitations within a given fermion number sector of the Fock space is in
principle possible in any dimensions. But, another interesting low dimensional
"accident", the bosonized effective theory \cite{bosope,bosopk,bospi} is 
local in dimension. The bosonized Luttinger model consists of free bosonic 
excitations which stand for the asymptotical, multi particle-hole states of 
the fermionic representation \cite{bosope,bosopk,mltmsol}. 
The most striking phenomenon
of this chiral model is the loss of Fermi-liquid properties, the
continuity of momentum distributions at the Fermi points
at vanishing temperature. In the absence of a transverse direction to 
the motion the slightest interactions destabilize the particle-like
asymptotical states which always propagate with the same speed.
The asymptotical states of the theory, if they exist, 
must correspond either to massive particles or to non-interacting
quasi-particles.

The non-interacting nature of the
asymptotical sector seems to be consistent with the vanishing of the 
beta functions, confirmed in a number ways, independently of the bosonization.
One chain of arguments is based on the application of Ward identities 
which are used to show a perfect cancellation between the self energy 
and the vertex corrections and to render the simple RPA exact 
\cite{mdcea,mdceb,mdck}.
The Wilsonian blocking employed in momentum space for the coupling constant 
of the bare action produces vanishing beta function
\cite{shankar,schultz}, too.

It is worthwhile mentioning a superficial analogy with the
confinement of color charges in QCD. In fact, the non-propagating
nature of the elementary fermionic excitations in one dimension
is equivalent to the cancellation of the residue of the particle pole
in the Green functions and the scattering amplitude. These are in turn
the hallmark of confinement in hadron physics, the absence of
colored asymptotical states. But the closer look betrays
that the two phenomena are actually different.
In fact, the bare particles of the one-dimensional world
remain non-propagating for either sign of the coupling
constant, i.e. both for attractive and for repulsive interactions,
as a result of the kinematical origin of the effect. On the contrary,
the attractive color forces are crucial in removing the colored
asymptotical states in QCD.

Nevertheless, the superficial analogy between these cases raises
other puzzles. The quark confinement is already
a rather spectacular phenomenon because it changes the degrees of
freedom with the scale of observation. But the strong correlations
between far separated quarks is reflected in the strong interactions
among the quasiparticles, the hadrons. What happens in the bosonized
model is even more surprising: How could the complete screening 
between "hadrons" take place in contrast to the strong, long-range 
correlations between the elementary constituents? 
Another question comes from the fact that
classically scale invariant, renormalizable models display
dimensional transmutation \cite{dimtr}. It arises when a (partial)resummation
of the perturbation expansion produces a combination of the cutoff $\Lambda$ 
and the dimensionless bare couplings in observables which remains finite in 
the limit $\Lambda\to\infty$. This must happen unless the IR scaling laws
remain the same as in the UV regime. The non-interacting, massless
particle-hole quasiparticles posses no scale, any
scale generated would indicate finite interaction strength. 
Why there is no dimensional transmutation, or different scaling
regimes in the Luttinger model? No other classically scale invariant, 
renormalizable model shows such a simple structure.

These questions lead to a one-loop renormalization group study
of the Luttinger model \cite{djl} which gave non-trivial beta functions,
strongly coupled dynamics and singular momentum dependence in the effective 
interaction strength around the Fermi point. Furthermore,
there were UV Landau poles in certain kinematical regimes of the
effective coupling strengths, suggesting problems with renormalizability.
This introduces the question whether the Luttinger model is indeed
renormalizable.

We argue below that renormalization schemes based on normal ordering
are rather fragile in the sense that cutoff independent, finite
physical effects, such as dimensional transmutation and Landau poles,
may arise from parts of the renormalized 
action which are vanishing when the cutoff is removed. In order
to see if such phenomena are indeed present in the
Luttinger model we turn to another powerful method, the 
functional renormalization group in the internal space \cite{int}. 

In the usual renormalization group
strategy the UV cutoff is lowered and the dynamics of the
degrees of freedom, eliminated in this manner, is taken into
account by adjusting the parameters of an effective bare action.
On contrary, the fluctuations are turned on gradually in the
internal space scheme and their effects are accumulated
by the evolution equation for the effective action \cite{rev}.
The advantage of this method is that it induces weaker
non-local artifacts as blocking in momentum space and it gives
access to effective interaction strengths off the Fermi points.
These features are essential for the  Luttinger model where 
eventual strong interactions appear as a singularity of the 
effective interaction strengths in momentum space in the 
vicinity of the Fermi points.

The only approximation needed in the functional renormalization group
method is the restriction of the evolution equation onto a functional space 
with manageable size. The perturbation expansion
and the functional renormalization group based solutions of
a model can be contrasted by stating that one resums infinitely
many orders generated by few relevant vertices in the former
and uses one-loop expressions but infinitely many
effective vertices in the latter procedure. We made a modest step
in this work in extending the computation towards more coupling
constants and considered the Luttinger model with two running
coupling constants whose flow is obtained in the one-loop level.

Our results confirm the absence of evolution of the coupling constant
when it is considered just at the Fermi points. But as soon as one comes 
off the Fermi points even by an infinitesimal amount our renormalization 
group scheme produces a non-trivial evolution which renders the
theory strongly coupled in the vicinity of the Fermi point for arbitrary
bare parameters. The cutoff generates Landau poles which seems to block 
the way of the removal of the cutoff. These results contradict the
picture emerging from the normal ordering solutions and suggest
the survival of cutoff effects.

The plan of the paper is the following. The model and its generating functionals
are introduced in Section II. A peculiar feature of the normal ordering
schemes is discussed in Section III. The internal space renormalization group
is briefly described and applied to the Luttinger model in Section IV.
The projection of the evolution of the equation into a suitable chosen
functional subspace is discussed in Section V. The projected evolution
equation for the effective action is given in Section VI.
The simplest application of the evolution equation is presented in
Section VII where the effective coupling strength is chosen at the Fermi points
and the vanishing beta function is recovered. Section VIII contains the
main results of this work, the integration of the beta functions
for coupling strength off the Fermi points. Our results are summarized
in Section IX. Two appendices facilitate the reading of the
paper. The first sums up the notation and conventions used in the paper
and the second contains few details of the computation of the one-loop
integrals of the evolution equation.

\section{Luttinger model}\label{lutmo}
The Luttinger model and its generating functional for the connected and the
one-particle irreducible (1PI) Green functions are introduced in this section.
The model is defined by the action 
$S[\psid,\psi]=S_0[\psid,\psi]+S_i[\psid,\psi]$, with
\bea\label{bact}
S_0[\psid,\psi]&=&\sum_\sigma\int_x\psid_{\sigma;x}
[-\partial_t+\epsilon_\sigma(-i\partial_\v{x})]\psi_{\sigma;x}\nn
S_i[\psid,\psi]&=&\sum_{\sigma_1,\sigma_2,\sigma_3,\sigma_4}
\int_{x_1,y_1,x_2,y_2}g^{\sigma_1,\sigma_2,\sigma_3,\sigma_4}_{x_1,y_1,x_2,y_2}
\psid_{\sigma_1,x_1}\psi_{\sigma_2,y_1}\psid_{\sigma_3,x_2}\psi_{\sigma_4,y_2},
\eea
where the field operator $\psi_{\sigma;x}$ stands for left or right
moving fermions with $\sigma=-1$ or $+1$, respectively and
\be
\epsilon_\sigma(\v{p})=\sigma\v{p}-k_F.
\ee
See Appendix \ref{not} for notations and conventions.
The separate conservation of number of left and right movers is achieved
by allowing the interactions $g^{+,+,+,+}_{x_1,y_1,x_2,y_2}$,
$g^{-,-,-,-}_{x_1,y_1,x_2,y_2}$, $g^{+,+,-,-}_{x_1,y_1,x_2,y_2}$, and
$g^{-,-,+,+}_{x_1,y_1,x_2,y_2}$ only to be non-vanishing.
Symmetry under space inversion requires
$g^{+,+,+,+}_{x_1,y_1,x_2,y_2}=g^{-,-,-,-}_{x_1,y_1,x_2,y_2}$ and
$g^{+,+,-,-}_{x_1,y_1,x_2,y_2}=g^{-,-,+,+}_{x_1,y_1,x_2,y_2}$.

In order to determine the space-dependence of the coupling constants
one introduces the Fourier transforms
\be
\psi_{\sigma;x}=\int_p\Theta(\sigma\v{p})e^{ipx}\psi_{\sigma;p},~~~
\psid_{\sigma;x}=\int_p\Theta(\sigma\v{p})e^{-ipx}\psid_{\sigma;p}
\ee
and
\be
g^{\sigma_1,\sigma_2,\sigma_3,\sigma_4}_{x_1,y_1,x_2,y_2}
=\int_{p_1,p_2,p_3,p_4}e^{ip_1x_1-ip_2y_1+ip_3x_2-iy_2p_4}
\delta_{p_1-p_2+p_3-p_4;0}\hat{g}^{\sigma_1,\sigma_2,\sigma_3,\sigma_4}_{p_1,p_2,p_3,p_4}.
\ee
The coupling constants which order the Grassmanian fields in the path integral
according to Eq. \eq{bact} are given by
\bea
\hat g^{-,-,+,+}_{p_1,p_2,p_3,p_4}&=&e^{i\eta(2p_1^0-p_2^0-p_3^0+2p_4^0)}g^\times\nn
\hat g^{+,+,+,+}_{p_1,p_2,p_3,p_4}&=&e^{i\eta(2p_1^0-p_2^0-p_3^0+2p_4^0)}g^\sp
\eea
where $g^\times$ and $g^\sp$ are constants and $\eta=0^+$.

The generating functional $W[j,j^\dagger]$ for the connected Green functions is
given by
\be\label{gfcongrf}
e^{W[j,j^\dagger]}=\int\cd{\psi}\cd{\psid}e^{-S[\psid,\psi]
+\sum_\sigma(\psid_\sigma\cdot j_\sigma + j^\dagger_\sigma\cdot\psi_\sigma)}.
\ee

The effective action, defined by the Legendre transformation,
\be
\Gamma[\psid,\psi]=-W[j,j^\dagger]+\sum_\sigma
\pa{\Psid_\sigma\cdot j_\sigma + j^\dagger_\sigma\cdot\Psi_\sigma}
\ee
with
\bea
\Psid_{\sigma;x}&=&-\fd{W}{j_{\sigma;x}}\nn
\Psi_{\sigma;x}&=&\fd{W}{j^\dagger_{\sigma;x}}
\eea
generates the 1PI vertex functions by its functional Taylor expansion.

The generating functional \eq{gfcongrf} needs regulator. For the sake
of a better insight into the cutoff effects let us mention four
easily implementable regulators. (i) The simplest regulator is a sharp cutoff
in momentum space, the restriction of the Fourier modes of the field
configuration to $|\v{p}-\sigma k_F|<\Lambda$. (ii) We can construct
a slightly different sharp cutoff regulators by placing the system
onto space-time lattice. The most important change is the bending 
of the dispersion relations into smooth, periodic functions. (iii)
We shall need smooth cutoff which is easiest to reach by smearing
the local terms in the interaction Lagrangian. One possibility is
to introduce the smeared field
\be
\psi_{B~\sigma,x}=\int_y\rho_{x-y}\psi_{\sigma,y}
\ee
and use the action $S[\psid,\psi]=S_0[\psid,\psi]+S_i[\psid_B,\psi_B]$
in Eq. \eq{gfcongrf}. A natural choice for smearing is
\be
\rho_x=\frac{\Lambda^2}{\pi}e^{-x^2\Lambda^2}
\ee
which amounts to the use of 
\be
\psi_{B~\sigma,p}=\rho_p\psi_{\sigma,p}
\ee
with
\be
\rho_p=e^{-p^2/4\Lambda^2}.
\ee
We note that the smearing function $\rho_p=\rho_\Lambda(\v{p})$, with
\be
\rho_\Lambda(\v{p})=\begin{cases}1&|\v{p}|\le\Lambda\cr0&|\v{p}|>\Lambda
\end{cases}
\ee
brings us back to sharp momentum cutoff. (iv) Another smooth
cutoff which preserves the local form of the current in the interactions
and splits the two currents only is obtained by using the interaction
\be
S_{iB}[\psid,\psi]=\sum_{\sigma_1,\sigma_2,\sigma_3,\sigma_4}
\int_{x_1,y_1,x_2,y_2,z}\rho_z
g^{\sigma_1,\sigma_2,\sigma_3,\sigma_4}_{x_1,y_1,x_2,y_2}
\psid_{\sigma_1,x_1+z}\psi_{\sigma_2,y_1+z}\psid_{\sigma_3,x_2}\psi_{\sigma_4,y_2}.
\ee
in Eq. \eq{gfcongrf}. Each regularization admits bare Ward-identities.

\section{BPHZ scheme and bosonization}
Most of the studies of the Luttinger model was carried out
in an elegant subtraction scheme in the Hamiltonian formalism
which is based on normal ordering. We highlight in this section
the peculiarity of the results by showing that the normal ordering 
is actually very similar to the BPHZ subtraction scheme. 

The BPHZ subtraction scheme is based on the observation  \cite{bopa,hep}
that the primitive divergences of an irreducible Green function appear 
in the first few order of the Taylor expansion in the external momenta
around zero. These divergences are subtracted by the introduction of local
counterterms, given just by these few divergent terms. This procedure
can be organized in such a manner that the subtraction is carried out
on the level of the integrand of the loop integrals \cite{zim}. The
resulting scheme is remarkable because it has no divergences whatsoever,
the loop integrals are manifestly finite \cite{hzz}.

The bosonization in 1+1 dimensions has been thoroughly discussed 
in a number of papers \cite{lut,bosope,bosopk,bospi} and we are content
to make few remarks about the cutoff effects by making the regularization 
explicit. The bosonization starts with the explicit construction of
the correspondence between the fermionic and the bosonic field
operators and continues with the demonstartion that the Hamiltonian of 
the Luttinger model is of massless, 
noninteracting boson fields \cite{lut,bosope}. We shall consider the 
the verification of the bosonic commutators in details.

The field operator corresponding to a chiral fermion in a box 
of size $L$ and periodic boundary conditions is given by
\be\label{chfer}
\psi(x)=\frac{1}{L}\sum_ke^{ikx}c_k,
\ee
$k=2\pi n_k/L$, where $n_k$ is integer. These operators
obey the set of canonical anticommutation relation whose
only non-vanishing member in momentum space is 
$\{c_k,c^\dagger_{k'}\}=\delta_{k,k'}$. 
The boson operators are introduced by the equation
\be
b_q=-\frac{i}{\sqrt{n_q}}\sum_kc^\dagger_{k-q}c_k
\ee
for $q>0$ and
\be
\phi(x)=\frac{1}{L}\sum_{q>0}e^{iqx}b_k.
\ee
The canonical bosonic commutators for $b_q$ and $b^\dagger_q$
follow in a trivial manner except the relation for
\be\label{ntrcr}
[b_q,b^\dagger_q]=\frac{1}{n_q}\sum_kc^\dagger_kc_k
-\frac{1}{n_q}\sum_kc^\dagger_{k-q}c_{k-q}.
\ee
The diagonal matrix elements of this equation contains the difference of 
two divergent quantities and needs a cutoff.  The regulator plays two 
roles simultaneously: It (i) makes the expressions finite by suppressing 
the short distance modes and, and (ii) defines an order of the 
summation/integration. The latter is rarely needed because the counterterms,
pulled into the loop summation/integration render the convergence absolute and 
hence independent of the order. But there might observables with 
sums/integrals which are finite but non-absolutely convergent. These have no
counterterms but still need a definite and consistent prescription
in order to render the theory well defined. These observables are called
anomalous.

In lattice regularization the 
Brillouin zone is periodic and the shift $k\to k+q$ in the first term
is allowed and renders the commutator vanishing. The same argument
applies for sharp momentum cutoff, as well \cite{lut}. 

But the regularization scheme where smooth cutoff is combined by normal 
ordering gives different result \cite{bosope}. We shall use the regulator
(iii) introduced in Section \ref{lutmo} which yields
\be
b_{Bq}=-\frac{i}{\sqrt{n_q}}\sum_k\rho^*_{k-q}\rho_kc^\dagger_{k-q}c_k
\ee
as bare operators for the interaction Lagrangian.
We define the fermionic vacuum by $c_k|0\ra=c^\dagger_{k'}|0\ra=0$ 
for $k,-k'>0$ and the corresponding normal ordering of an operator $O$ is
\be
:O:=O-\la0|O|0\ra.
\ee
The regulated Eq. \eq{ntrcr} now reads as
\bea\label{ntrcrs}
[b_{Bq},b^\dagger_{Bq}]&=&\frac{1}{n_q}\sum_k|\rho_{k-q}|^2c^\dagger_kc_k
-\frac{1}{n_q}\sum_k|\rho_k|^2c^\dagger_{k-q}c_{k-q}\\
&=&\frac{1}{n_q}\sum_k|\rho_{k-q}|^2:c^\dagger_kc_k:
-\frac{1}{n_q}\sum_k|\rho_k|^2:c^\dagger_{k-q}c_{k-q}:
+\frac{1}{n_q}\sum_k|\rho_k|^2\la0|c^\dagger_{k-q}c_{k-q}|0\ra
-\frac{1}{n_q}\sum_k|\rho_{k-q}|^2\la0|c^\dagger_kc_k|0\ra.\nonumber
\eea
One can show that any matrix element of the sums is absolute
convergent and the reordering $k\to k+q$ is allowed, giving
\be\label{ntrcrss}
[b_{Bq},b^\dagger_{Bq}]=\openone(1+C_q)+D_q
\ee
where
\bea
C_q&=&\frac{1}{n_q}\sum_{k<0}(|\rho_k|^2-|\rho_{k-q}|^2)
+\frac{1}{n_q}\sum_{k=0}^q(|\rho_k|^2-1)\nn
D_q&=&\frac{1}{n_q}\sum_k(|\rho_{k-q}|^2-|\rho_{k+q}|^2):c^\dagger_kc_k:
\eea
In the works using bosonization the usual regulator is (iv) of 
Section \ref{lutmo} which leads to Eq. \eq{ntrcrs} with $\rho_k=1$.
The sum with the normal ordered product gives absolute convergent
matrix elements but the last two sums diverge and renders this cutoff
unacceptable.

In order to understand better the problem of the regularization (iv) 
let us point out the similarity of the normal ordered scheme and the 
BPHZ method. One subtraction is sufficient in 
1+1 dimensions and only its value differs in the normal ordered and the 
BPHZ schemes. The normal ordering shares the advantages of the BPHZ scheme,
namely simplicity, the disappearance of divergences. But this does not mean 
that these schemes need no regulator whatsoever. The following two 
circumstances make the regulator unavoidable. 

If there are non-absolutely convergent but finite loop summations/integration, 
anomalous observables, in the theory then regulator is needed in order to 
define their value in a consistent manner. The consistency is important,
eg. one could have chosen any order in the second summation of the
last line in Eq. \eq{ntrcrs} with $\rho_k=1$ which yields finite result.
But it is not obvious how to define sums which appear in other loops.
One needs a consistent regulator which settle this issue on the level
of the bare Hamiltonian and not of particular expressions. 

The other instance which requires regulatrization is when the cutoff manages 
to generate a finite scale. An example is the dimensional 
transmutation. The only way the dimensional transmutation can appear 
in the BPHZ scheme is through the finite cutoff-dependence which is vanishing 
in the limit $\Lambda\to\infty$ because all divergent cutoff dependence
which generates the new scale in other schemes are now subtracted.
Therefore the finite expressions of the cutoff which tend to
zero in the limit $\Lambda\to\infty$ must be retained
before resumming the perturbation expansion in order no tot miss
dimensional transmutation. There are simpler
or more complicate schemes, where the same renormalized physics is 
recovered in a straightforward or more involved manner. The BPHZ scheme
which is very powerful for massive models without dimensional
transmutation but becomes highly nontrivial for massless models.
This happens because its renormalized parameters are not closely related to the
true effective coupling constants and finite, physical effects
(eg. dimensional transmutation) arise from infinitesimal small
modifications of the BPHZ renormalized dynamics. As far as the BPHZ
is concerned, one can save it with the expense of some complications,
by expanding the irreducible Green functions around non-vanishing
energy-momenta, but the normal ordering can not be rescued
in the absence of any scale to rely upon.

Another case when the regulator is needed is when the cutoff
can not be moved beyond a certain threshold. This can easily happen
in a perturbatively renormalizable theory, where the divergences
are eliminated order by order in the perturbation expansion.
In fact, the renormalization condition which fixes the theory
is given as a set of equations, containing certain experimentally
observed values on one side and the corresponding expressions on
the other side. These expressions contain the cutoff and the bare 
parameters in a highly non-linear manner and it may happen that
they have no solution with finite values of the bare parameters 
beyond certain scale, usually called Landau pole scale, for the cutoff.

After these general remark we now return to the discussion of Eq. \eq{ntrcrss}.
The commutation relation of Ref. \cite{bosope} is obtained from Eq.
\eq{ntrcrss} by setting $C_q=D_q=0$, quantities of $\ord{q^2/\Lambda^2}$.
After this simplification the bosonization is quickly completed by
expressing the normal ordered fermionic Hamiltonian by means of
the boson field and by performing a Bogoliubov transformation.
The result is a normal ordered Hamiltonian for a massless,
non-interactive boson field.

Our point is that the regulator (iv) regulates all Feynman graphs 
contributing to Green functions because it suppresses interactions beyond 
the cutoff scale. But it does $not$ remove divergences, such as in 
Eq. \eq{ntrcrss}, from the unperturbed dynamics even if the regulator
is amended by normal ordering. Normal ordering is a reordering
of the loop contributions only and not regulator.
The control of $all$ divergences actually leads to
the $\ord{q^2/\Lambda^2}$ terms in Eq. \eq{ntrcrss}. They are certainly
negligible as far as the canonical commutation relations are concerned
at face value in the renormalized theory but they modify the bosonic 
Hamiltonian which is based on these relations and generate some 
$\ord{\Lambda^{-2}}$ interactions. These interactions which 
are dropped by the formal limit $\Lambda\to\infty$
in the normal ordered scheme may generate finite effects
via dimensional transmutations or Landau poles. 

We shall, in this work, look for cutoff effects in the theory
defined by sharp momentum space cutoff in order to see if the 
omission of the $\ord{q^2/\Lambda^2}$ terms in Eq. \eq{ntrcrss}
is indeed justified. This choice of regulator brings us back to the 
discussion of different regularizations after Eq. \eq{ntrcr}. Have we not 
already lost bosonization with all of its consequences when sharp momentum 
cutoff is used? 

The fact that different, but acceptable regularizations produce 
different result for a finite value of the cutoff is not surprising. 
All what one expects from renormalizable theories is that different
cutoff theories converge to the same renormalized one when the
cutoffs are removed. The usual argument based on universality
assures that there is a unique Luttinger model, as long as its theory is
renormalizable, whatever acceptable regulator is used. It might
naturally happen that certain effects are more or less easily
identifiable when a given regulator is used but the numerical
value of the observables must agree. For example, there is a well known
no-go theorem \cite{nogo} that chiral fermions, as in Eq. \eq{chfer}, 
can not be accommodated within the framework of the usual quantum field 
theories when lattice regulator is used. But rather involved
constructions which can be viewed as particular regularization schemes
and violate the conditions of the no-go theorem can realize chiral 
lattice fermions in particular limits. In a similar manner, we may
regard the fragility of bosonization as an indication that this
scheme concentrates a large amount of informations in the 
$\ord{\Lambda^{-2}}$ part of the renormalized action. In other schemes
these informations spread over the finite part of the action. 
If they play appreciable role in the dynamics then bosonized models 
require an unusually careful treatment.

\section{Renormalization group in the internal space}
The traditional renormalization group method \cite{wilson} consists of the
successive elimination of the modes. There are two possible strategies
in determining the order of the elimination. The conventional
way is to eliminate modes at the running UV cutoff $\Lambda$ which moves in
the IR direction. One can imagine two different ways to justify this procedure.
(i) One starts with the unavoidable non-local effects of quantum field theories.
It is well known that a theory is always non-local at distances
which are shorter than the UV cutoff scale. These involved and dangerous
correlations are turned on gradually during the blocking if the UV cutoff
moves from the UV to the IR direction. (ii) Another strategy to
fix the order of the modes to be eliminated is that
the more perturbative modes should be eliminated by means of a simpler
effective theory. Hence one should tackle with the more nonperturbative
modes by means of a more sophisticated, "dressed" effective theory which is
obtained after having already eliminated a large number of modes. The
kinetic energy usually renders the short distance modes perturbative
therefore the modes are eliminated in decreasing order of their
momentum, their scale parameter in the space-time, called external space.
This latter corresponds to the convention of considering the
field variables $\phi(x)$ as a mapping from the external into
the internal space.

The caveat in the reasoning (i) is that the non-local effects
usually reach well beyond the naive length scales $1/\Lambda$.
In fact, let us consider a smooth momentum space
cutoff which leaves the modes with momentum $p<\Lambda$ unchanged
and suppresses modes with $p>c\Lambda$ ($c>1$). The details
of the suppression mechanism appears in effective vertices acting
on distance scales $[1/c\Lambda,1/\Lambda]$. These details
of the cutoff are arbitrary and non-physical therefor these non-local effects
are non-physical, too. In particular, sharp momentum cutoff always
generates infinitely long range correlations. Their strength
is vanishing as $\Lambda\to\infty$ in a renormalizable theory but
they are present in effective theories where $\Lambda$ is finite.

One might argue that the subsequent steps of the lowering of
the cutoff always cancel the non-local contributions of the previous step
and replaces them by the non-local contribution corresponding
to the lowered value of the cutoff. But any truncation of the
evolution equation raises the issue of approximate cancellation 
and forces us to search of alternative blocking schemes which generate 
as weak non-local effects as possible.

Such an alternative scheme can be found by generalizing the reasoning (ii).
The point is that there are two ways a mode may become perturbative.
One, covered by reasoning (ii), is that the corresponding
action is dominated by the kinetic term. But another
possibility is that its amplitude is suppressed by the potential energy.
This second way a mode may turn out to be perturbative is exploited
when the order of the elimination proceeds according to the amplitude
of the modes, the scale in the internal space \cite{int}.
One starts with modes of small amplitudes leaving the
large amplitude modes at the end. This procedure amounts to the realization
of the Callan-Symanzik proposal \cite{calsym} when the amplitude of the
modes are controlled by the momentum independent mass term in a 
relativistic theory.

There are cases when the momentum independent suppression
is not available, when symmetry excludes mass like in gauge theories.
A possible solution to this problem can be found by generalizing
this scheme and taking any continuous parameter in the dynamics as
a control parameter which (a) corresponds to a quadratic expression of
the fields and (b) the theory is soluble for certain limit of this parameter.
The condition (a) may require the realization of this program within the
framework of the effective theory for composite operators, such as $\phi^2$ in
the $\phi^4$ scalar model \cite{int} or fermion bilinears \cite{ps,aps,loc,hoen}.

What kind of non-local effects are generated by renormalizing in the internal space? 
This is an essential question for the Luttinger model where the collinear IR 
divergences are expected to plague the perturbation expansion and to generate
long range effective interactions in terms of the elementary
fermionic excitations, as mentioned in the Introduction.

The Callan-Symanzik scheme in which the momentum
independent, relativistic mass term is used to control the
elimination of the modes \cite{int} appears to be a scheme with a
minimal amount of non-locality. When the mass term is
forbidden, as in the Luttinger model, then we may use the
coupling constant as a control parameter. More precisely,
we rescale the coupling constant by a factor $\lambda^2$ which will
be evolved from 0 to 1. This scheme will yield a gradual turning
on the "dangerous" part of the quantum fluctuations which are not
factorizable according to the Wick-theorem.
The only non-local feature of this suppression scheme comes from the
energy-momentum dependence of the interactions,
needed for the separation of the left and right modes. But this is
a $\lambda$-independent real physical effect and is a necessary ingredient.

Another advantageous feature of this scheme when applied for fermionic
systems can be seen in the following manner.
In fermionic models at finite density the dangerous, soft modes are
around the Fermi surface and the blocking should zoom in this area
at the final stage of the evolution. This usually requires a rather
involved geometric construction which can easily be avoided
by controlling the interaction strength. A rescaling of the fermionic
fields by $\lambda^{-1/2}$ removes the factor $\lambda$ from the
usual quartic interaction term and generates the coefficient
$1/\lambda$ in front of the quadratic part of the action, i.e.
multiplies the propagators by $\lambda$. One can see again that the
momentum dependence brought by this control parameter is dictated by
the the true, $\lambda$-independent dynamics, the kinetic energy,
instead of an artificial device and it may build up the eventually strongly
coupled dynamics around the Fermi surface.

Beyond the issue of preserving locality better, the most important difference
between the external and internal space schemes from the point of view
of fermionic system is the possibility in the latter case to obtain
effective interaction strengths off the Fermi surface. In fact, the
external space blocking methods zoom into the Fermi surface at the
end point of the evolution only. As a result, any off-Fermi surface region is
handled without the dynamical contributions of modes between the
region in question and the Fermi surface. This happens because these
modes appear at a later stage of the evolution only when the effective coupling
strengths off the Fermi surface have already been fixed. On the contrary,
the internal space scheme blocking has no impact on the external space
and leaves the possibility of tracing the evolution of the effective
coupling strengths at in an arbitrary kinematical region.

The realization of the internal space blocking scheme starts with the extension
\be
S'_\lambda[\psid,\psi]=\sum_\sigma\int_x\psid_{\sigma;x}
[-\partial_t+\epsilon_\sigma(-i\partial_\v{x})]\psi_{\sigma;x}
+\lambda^2\sum_{\sigma_1,\sigma_2,\sigma_3,\sigma_4}
\int_{x_1,y_1,x_2,y_2}g^{\sigma_1,\sigma_2,\sigma_3,\sigma_4}_{x_1,y_1,x_2,y_2}
\psid_{\sigma_1,x_1}\psi_{\sigma_2,y_1}\psid_{\sigma_3,x_2}\psi_{\sigma_4,y_2}
\ee
of the action \eq{bact} which can be written after the rescaling
$\psi\to\lambda^{1/2}\psi$, $\psid\to\lambda^{1/2}\psid$ as
\be\label{ract}
S_\lambda[\psid,\psi]=\frac{1}{\lambda}\sum_\sigma\int_x\psid_{\sigma;x}
[-\partial_t+\epsilon_\sigma(-i\partial_\v{x})]\psi_{\sigma;x}
+\sum_{\sigma_1,\sigma_2,\sigma_3,\sigma_4}
\int_{x_1,y_1,x_2,y_2}g^{\sigma_1,\sigma_2,\sigma_3,\sigma_4}_{x_1,y_1,x_2,y_2}
\psid_{\sigma_1,x_1}\psi_{\sigma_2,y_1}\psid_{\sigma_3,x_2}\psi_{\sigma_4,y_2}
\ee

The evolution equation for the generating functional is obtained by simply
applying the derivative with respect to the control parameter $\lambda$
on Eq. \eq{gfcongrf} where the action is given by Eq. \eq{ract},
\bea
\partial_\lambda W[j_\sigma,j^\dagger_\sigma]
&=&\frac{1}{\lambda^2}e^{-W[j^\dagger,j]}\int\cd{\psi}\cd{\psid}
e^{-S_\lambda[\psid,\psi]+\sum_\sigma(\psid_\sigma\cdot j_\sigma+j_\sigma^\dagger\cdot\psi_\sigma)}\nn
&&\times\int_x\sum_\sigma\psid_{\sigma;x}[-\partial_t+\epsilon_\sigma(-i\partial_x)]\psi_{\sigma;x}\nn
&=&\frac{1}{\lambda^2}\sum_\sigma\tr\left[\left(W^{(2)}_{j_\sigma^\dagger,j_\sigma}
+W_{j_\sigma^\dagger}W_{j_\sigma}\right)
\cdot[-\partial_t+\epsilon_\sigma(-i\partial_x)]\right].
\eea
The truncation of this functional differential equation is rather
difficult because the functional $W[j_\sigma,j^\dagger_\sigma]$ is 
too complicated due to strong non-local features. The 1PI vertex
functions are supposed to be more local than the connected
Green functions therefore we express this equation in terms of
the effective action. We introduce the notation
\be
F^{(2)}_{\phi,\chi}=\fdd{F[\phi,\chi]}{\phi}{\chi},
\ee
use the relation
\bea
&&\int_y\begin{pmatrix}
\Gamma^{(2)}_{\Psid_{+,z},\Psi_{+;y}}&\Gamma^{(2)}_{\Psid_{+,z},\Psi_{-;y}}
&\Gamma^{(2)}_{\Psid_{+,z},\Psid_{+;y}}&\Gamma^{(2)}_{\Psid_{+,z},\Psid_{-;y}}\cr
\Gamma^{(2)}_{\Psid_{-,z},\Psi_{+;y}}&\Gamma^{(2)}_{\Psid_{-,z},\Psi_{-;y}}
&\Gamma^{(2)}_{\Psid_{-,z},\Psid_{+;y}}&\Gamma^{(2)}_{\Psid_{-,z},\Psid_{-;y}}\cr
\Gamma^{(2)}_{\Psi_{+,z},\Psi_{+;y}}&\Gamma^{(2)}_{\Psi_{+,z},\Psi_{-;y}}
&\Gamma^{(2)}_{\Psi_{+,z},\Psid_{+;y}}&\Gamma^{(2)}_{\Psi_{+,z},\Psid_{-;y}}\cr
\Gamma^{(2)}_{\Psi_{-,z},\Psi_{+;y}}&\Gamma^{(2)}_{\Psi_{-,z},\Psi_{-;y}}
&\Gamma^{(2)}_{\Psi_{-,z},\Psid_{+;y}}&\Gamma^{(2)}_{\Psi_{-,z},\Psid_{-;y}}
\end{pmatrix}\begin{pmatrix}
-W^{(2)}_{j^\dagger_{+,y},j_{+;x}} & -W^{(2)}_{j^\dagger_{+,y},j_{-;x}}
&W^{(2)}_{j^\dagger_{+,y},j^\dagger_{+;x}}&W^{(2)}_{j^\dagger_{+,y},j^\dagger_{-;x}}\cr
-W^{(2)}_{j^\dagger_{-,y},j_{+;x}}&-W^{(2)}_{j^\dagger_{-,y},j_{-;x}}
&W^{(2)}_{j^\dagger_{-,y},j^\dagger_{+;x}}&W^{(2)}_{j^\dagger_{-,y},j^\dagger_{-;x}}\cr
W^{(2)}_{j_{+,y},j_{+;x}}&W^{(2)}_{j_{+,y},j_{-;x}}
&-W^{(2)}_{j_{+,y},j^\dagger_{+;x}}&-W^{(2)}_{j_{+,y},j^\dagger_{-;x}}\cr
W^{(2)}_{j_{-,y},j_{+;x}}&W^{(2)}_{j_{-,y},j_{-;x}}
&-W^{(2)}_{j_{-,y},j^\dagger_{+;x}}&-W^{(2)}_{j_{-,y},j^\dagger_{-;x}}
\end{pmatrix}\nn
&&~~~=-\delta_{z,x}\begin{pmatrix}1&0&0&0\cr0&1&0&0\cr0&0&1&0\cr0&0&0&1\end{pmatrix},
\eea
and find
\bea
\partial_\lambda\Gamma[\Psid,\Psi]
&=&-\partial_\lambda W[j_\sigma,j^\dagger_\sigma]\nn
&=&\frac{1}{\lambda^2}\sum_\sigma\tr\left[\left(\Gamma_{\Psid_\sigma,\Psi_\sigma}^{^{(2)}-1}
-\Psi_\sigma\Psid_\sigma\right)\cdot[-\partial_t+\epsilon_\sigma(-i\partial_x)]\right].
\eea

The solution will be sought by means of the functional form
\be
\Gamma[\Psid,\Psi]=\tilde\Gamma[\Psid,\Psi]+\frac{1}{\lambda}\sum_{\sigma=\pm1}\Psid_\sigma\cdot
[-\partial_t+\epsilon_\sigma(-i\partial_x)]\cdot\Psi_\sigma,
\ee
for which the evolution equation reads as
\bea\label{evoleq}
\partial_\lambda\tilde\Gamma[\Psid,\Psi]
&=&-\frac{1}{\lambda^2}\sum_\sigma\tr\left(\Gamma_{\Psid_\sigma,\Psi_\sigma}^{^{(2)}-1}
\cdot[-\partial_t+\epsilon_\sigma(-i\partial_x)]\right).
\eea

\section{Ansatz}
The general solution of the functional differential equation \eq{evoleq}
is beyond our capacities and we have arrived at the only approximation
one has to rely upon in this scheme, the truncation of the evolution equation.
By assuming an analytical dependence in the field variables we write
$\tilde\Gamma[\Psid,\Psi]$ in the form of a functional Taylor series
where arbitrary high powers of the fields appear and the
coefficient functions, the 1PI vertices, are arbitrary
functions of their space-time or energy-momentum variables. The $n$-th
order terms in this series represent $n$-body correlations and
can be shown to be suppressed in the framework of the Wilsonian
the external space renormalization group method in the limit
when the sharp momentum space cutoff used approaches zero \cite{mdck}.
Though this result seems to be somehow fragile, the nonlocal interactions
generated by the sharp cutoff imposed in the IR regime may substantially
change the kinetic energy and thus invalidate the argument, we shall
ignore the the terms $n\ge6$
for the sake of simplicity. But the collinear divergences will be
taken more seriously and an arbitrary space-time dependence will be
retained for the vertex functions. Our choice for the "potential"
$\tilde\Gamma[\Psid,\Psi]$ is
\bea\label{ans}
\tilde\Gamma[\Psid,\Psi]&=&\Gamma_0-\sum_\sigma\Psid_\sigma\cdot\Sigma_\sigma\cdot\Psi_\sigma
-\int_{x_1,y_1,x_2,y_2}\gamma^\times_{x_1,y_1,x_2,y_2}\Psid_{-;x_1}\Psi_{-;y_1}
\Psid_{+;x_2}\Psi_{+;y_2}\nn
&&-\int_{x_1,y_1,x_2,y_2}\gamma^\sp_{x_1,y_1,x_2,y_2}(\Psid_{+;x_1}\Psi_{+;y_1}
\Psid_{+;x_2}\Psi_{+;y_2}+\Psid_{-;x_1}\Psi_{-;y_1}\Psid_{-;x_2}\Psi_{-;y_2}),
\eea
where the effective vertices obey the symmetries
$\gamma^\sp_{x_1,y_1,x_2,y_2}=\gamma^\sp_{x_2,y_2,x_1,y_1}=-\gamma^\sp_{x_1,y_2,x_2,y_1}$.

We shall be interested in the question if the theory is strongly coupled
around the Fermi points. It should be kept in mind that the eventual divergence
of the vertex functions at some points does not necessarily mean
that the model is strongly coupled in that kinematical regime. In fact,
any physically acceptable definition of the strength of the two-body
interaction involves a finite resolution in the energy-momentum space and
seeks an average, a representative value of the four-point vertex function in a
small but finite area. Such a construction can be approximated by the 
introduction of an appropriately chosen normalized, smooth
smearing function $\rho_{p,q,r,s}$ in the definition of the dimensionless
running coupling constant,
\be
\gamma=\int_{p,q,r,s}\delta_{p-q+r-s,0}\rho_{p,q,r,s}\hat\gamma_{p,q,r,s}
\ee
where $\hat\gamma_{p,q,r,s}$ is obtained from the vertex function by separating
off the trivial energy-momentum conservation,
\be
\gamma_{x_1,y_1,x_2,y_2}=\int_{p,q,r,s}e^{ipx_1-iqy_1+irx_2-isy_2}
\delta_{p-q+r-s,0}\hat\gamma_{p,q,r,s}.
\ee
Guided by such a construction we shall call the theory strongly coupled
at a certain kinematical point if the four point function $\hat\gamma_{p,q,r,s}$
either takes a uniformly large value around or displays a non-integrable singularity
at that point.

One can explore the weakly coupled regime of the theory in a consistent manner
with our truncation, \eq{ans}. But whenever the dimensionless running coupling
constants assume values which is not anymore small with respect to one
then our computation must stop.

\section{Evolution equation}
In order to arrive at the evolution equation for our ansatz we need
the inverse of the second functional derivative,
\be
\Gamma^{(2)}=\begin{pmatrix}
\Gamma^{(2)}_{\Psid_{+,x},\Psi_{+;y}}&\Gamma^{(2)}_{\Psid_{+,x},\Psi_{-;y}}
&\Gamma^{(2)}_{\Psid_{+,x},\Psid_{+;y}}&\Gamma^{(2)}_{\Psid_{+,x},\Psid_{-;y}}\cr
\Gamma^{(2)}_{\Psid_{-,x},\Psi_{+;y}}&\Gamma^{(2)}_{\Psid_{-,x},\Psi_{-;y}}
&\Gamma^{(2)}_{\Psid_{-,x},\Psid_{+;y}}&\Gamma^{(2)}_{\Psid_{-,x},\Psid_{-;y}}\cr
\Gamma^{(2)}_{\Psi_{+,x},\Psi_{+;y}}&\Gamma^{(2)}_{\Psi_{+,x},\Psi_{-;y}}
&\Gamma^{(2)}_{\Psi_{+,x},\Psid_{+;y}}&\Gamma^{(2)}_{\Psi_{+,x},\Psid_{-;y}}\cr
\Gamma^{(2)}_{\Psi_{-,x},\Psi_{+;y}}&\Gamma^{(2)}_{\Psi_{-,x},\Psi_{-;y}}
&\Gamma^{(2)}_{\Psi_{-,x},\Psid_{+;y}}&\Gamma^{(2)}_{\Psi_{-,x},\Psid_{-;y}}
\end{pmatrix},
\ee
computed in a manner which is consistent with the truncation.
For this end we write it in the form
\be
\Gamma^{(2)}=G^{-1}_0+\Delta^\times+\Delta^\sp
\ee
by means of the diagonal block matrix
\be
G^{-1}_{0~x,y}=\begin{pmatrix}
G_{+;x,y}^{-1} & 0 & 0 & 0 \\
0 & G_{-;x,y}^{-1} & 0 & 0 \\
0 & 0 & -G_{+;y,x}^{-1} & 0 \\
0 & 0 & 0 & -G_{-;y,x}^{-1}
\end{pmatrix}
\ee
composed of the free propagator
\be
G_\sigma^{-1}=\frac{1}{\lambda}[-\partial_t+\epsilon_\sigma(-i\partial_x)]+\Sigma_\sigma,
\ee
 the self energies
\be
\Delta^\times_{x,y}=\int_{z_1,z_2}\begin{pmatrix}
\gamma^\times_{z_1,z_2,x,y}\Psid_{-;z_1}\Psi_{-;z_2}
&-\gamma^\times_{z_1,y,x,z_2}\Psid_{-;z_1}\Psi_{+;z_2}
&0&\gamma^{\times}_{y,z_1,x,z_2}\Psi_{-;z_1}\Psi_{+;z_2}\cr
-\gamma^\times_{x,z_1,z_2,y}\Psid_{+;z_2}\Psi_{-;z_1}
&\gamma^\times_{x,y,z_1,z_2}\Psid_{+;z_1}\Psi_{+;z_2}
&\gamma^\times_{x,z_1,y,z_2}\Psi_{-;z_1}\Psi_{+;z_2}&0\cr
0&-\gamma^\times_{z_1,y,z_2,x}\Psid_{-;z_1}\Psid_{+;z_2}
&-\gamma^\times_{z_1,z_2,y,x}\Psid_{-;z_1}\Psi_{-;z_2}
&\gamma^\times_{y,z_1,z_2,x}\Psid_{+;z_2}\Psi_{-;z_1}\cr
-\gamma^\times_{z_1,x,z_2,y}\Psid_{-;z_1}\Psid_{+;z_2}&0
&\gamma^\times_{z_1,x,y,z_2}\Psid_{-;z_1}\Psi_{+;z_2}
&-\gamma^\times_{y,x,z_1,z_2}\Psid_{+;z_1}\Psi_{+;z_2}
\end{pmatrix},
\ee
and
\be
\Delta^\sp_{x,y}=\int_{z_1,z_2}\begin{pmatrix}
4\gamma^\sp_{x,y,z_1,z_2}\Psid_{+;z_1}\Psi_{+;z_2}&0&
2\gamma^\sp_{y,z_1,x,z_2}\Psi_{+;z_1}\Psi_{+;z_2}&0\cr
0&4\gamma^\sp_{x,y,z_1,z_2}\Psid_{-;z_1}\Psi_{-;z_2}&0&
2\gamma^\sp_{y,z_1,x,z_2}\Psi_{-;z_1}\Psi_{-;z_2}\cr
2\gamma^\sp_{z_1,y,z_2,x}\Psid_{+;z_1}\Psid_{+;z_2}&0&
-4\gamma^\sp_{y,x,z_1,z_2}\Psid_{+;z_1}\Psi_{+;z_2}&0\cr
0&2\gamma^\sp_{z_1,y,z_2,x}\Psid_{-;z_1}\Psid_{-;z_2}0&0&
-4\gamma^\sp_{y,x,z_1,z_2}\Psid_{-;z_1}\Psi_{-;z_2}
\end{pmatrix}.
\ee
The matrix elements of the second functional derivative in the desired approximation
are finally obtained by means of the Neumann expansion,
\be
(\Gamma^{(2)})^{-1}=G_0-G_0\cdot(\Delta^\times+\Delta^\sp)\cdot G_0
+G_0\cdot(\Delta^\times+\Delta^\sp)\cdot G_0\cdot
(\Delta^\times+\Delta^\sp)\cdot G_0+\cdots
\ee
which yields the evolution equation
\be
\partial_\lambda\tilde\Gamma=-\frac{1}{\lambda^2}\tr\left\{
[-\partial_t+\epsilon_\sigma(-i\partial_x)]
\cdot[G_0-G_0\cdot(\Delta^\times+\Delta^\sp)\cdot G_0
+G_0\cdot(\Delta^\times+\Delta^\sp)\cdot G_0\cdot
(\Delta^\times+\Delta^\sp)\cdot G_0]\right\}
\ee
for our ansatz. The first contribution arising from $G_0$ in the second
parenthesis on the right hand side is field independent and can be ignored.
The quadratic and quartic terms generate the evolution for the
self energy and the effective coupling constants, respectively, to be
considered below.

\subsection{Quadratic part}
The quadratic order of the left and the right hand sides of the evolution
equation are
\be
\int_{z_1,z_2}(\Psid_{+;z_1}\partial_\lambda\Sigma_{+,z_1,z_2}\Psi_{+,z_2}
+\Psid_{-;z_1}\partial_\lambda\Sigma_{-,z_1,z_2}\Psi_{-,z_2})
\ee
and
\bea
&&-\frac{1}{\lambda^2}\int_{z_1,z_2,z_3,z_4}[G_{+;z_4,z_1}
(\Delta^\times_{1,1}+\Delta^\sp_{1,1})_{z_1,z_2}
G_{+;z_2,z_3}[-\partial_t+\epsilon_+(-i\partial_x)]_{z_3,z_4}\nn
&&-\frac{1}{\lambda^2}\int_{z_1,z_2,z_3,z_4}[G_{-;z_4,z_1}
(\Delta^\times_{2,2}+\Delta^\sp_{2,2})_{z_1,z_2}
G_{-;z_2,z_3}[-\partial_t+\epsilon_-(-i\partial_x)]_{z_3,z_4}
\eea
respectively. By equating these expressions one finds the evolution
equations for the self energy,
\bea
\partial_\lambda\Sigma_{-,x,y}&=&-\frac{1}{\lambda^2}\int_{z_1,z_2,z_3,z_4}\{G_{+;z_4,z_1}
\gamma^\times_{x,y,z_1,z_2}G_{+;z_2,z_3}[-\partial_t+\epsilon_+(-i\partial_x)]_{z_3,z_4}\nn
&&+4G_{-;z_4,z_1}\gamma^\sp_{z_1,z_2,x,y}
G_{-;z_2,z_3}[-\partial_t+\epsilon_-(-i\partial_x)]_{z_3,z_4}\},\nn
\partial_\lambda\Sigma_{+,x,y}&=&-\frac{1}{\lambda^2}\int_{z_1,z_2,z_3,z_4}\{G_{-;z_4,z_1}
\gamma^\times_{x,y,z_1,z_2}G_{-;z_2,z_3}[-\partial_t+\epsilon_-(-i\partial_x)]_{z_3,z_4}\nn
&&+4G_{+;z_4,z_1}\gamma^\sp_{z_1,z_2,x,y}
G_{+;z_2,z_3}[-\partial_t+\epsilon_+(-i\partial_x)]_{z_3,z_4}\}.
\eea
The structure of these equations reflects the Callan-Symanzik strategy
in the framework of evolution in $\lambda$. In fact, the relation
\be
\partial'_\lambda G_\sigma=\partial'_\lambda\frac{1}
{\frac{1}{\lambda}[-\partial_t+\epsilon_\sigma(-i\partial_x)]+\Sigma_\sigma}
=-\frac{1}{\lambda^2}G_\sigma\cdot[-\partial_t+\epsilon_\sigma(-i\partial_x)]\cdot G_\sigma
\ee
where $\partial'_\lambda$ acts on the explicit, driving $\lambda$-dependence
produces
\bea
\partial_\lambda\Sigma_{-;x,y}&=&\partial'_\lambda\int_{z_1,z_2}
(\gamma^\times_{x,y,z_1,z_2}G_{+;z_2,z_1}+4\gamma^\sp_{z_1,z_2,x,y}G_{-;z_2,z_3}),\nn
\partial_\lambda\Sigma_{+;x,y}&=&\partial'_\lambda\int_{z_1,z_2}
(\gamma^\times_{x,y,z_1,z_2}G_{-;z_2,z_1}+4\gamma^\sp_{z_1,z_2,x,y}G_{+;z_2,z_3}),
\eea
the perturbative, one-loop Callan-Symanzik beta function for the propagator.
The Fourier transform of the self energy
\be
\Sigma_{\sigma;x,y}=\int_{p,q}e^{ixp+iyq}\delta_{p+q,0}\hat\Sigma_{\sigma;p,q},
\ee
allows us to write the evolution equation in a simpler form,
\be\label{quadr}
\partial_\lambda\Sigma_{\pm;p}=\partial'_\lambda\int_q
\frac{\hat\gamma^\times_{p,p,q,q}+4\hat\gamma^\sp_{p,p,q,q}}
{\frac{1}{\lambda}[-\partial_t+\epsilon_\mp(i\partial_x)]+\Sigma_\mp}.
\ee
It is worthwhile noting the appearance of different derivatives on the two sides
of these equations. In fact, had we found $\partial_\lambda$ on both sides
the solution would have produced the standard one-loop results for
the self energy. The derivative $\partial'_\lambda$ on the right hand
side makes sure that the effective vertices $\hat\gamma^\times$,
$\hat\gamma^\sp$ which in principle depend on $\lambda$ are nevertheless kept intact.
This procedure shows the essence of the infinitesimal blocking version
of the renormalization group strategy: the for an infinitesimal change
of the control parameter can safely be computed perturbatively by means
of the actual effective vertices and the nonperturbative resummation
arises from the repetition, i.e. the integration of this procedure.

\subsection{Quartic part}
The evolution equation gives in the quartic order
\bea
\lambda^2\partial_\lambda\gamma^\times_{x_1,y_1,x_2,y_2}
&=&-\int_{z_1,\ldots,z_6}\{G_{+;z_6,z_1}(-4\gamma^\times_{x_1,y_1,z_1,z_2}
G_{+;z_2,z_3}\tilde\gamma^\sp_{z_3,z_4,x_2,y_2}
-4\tilde\gamma^\sp_{z_1,z_2,x_2,y_2}G_{+;z_2,z_3}\gamma^\times_{x_1,y_1,z_3,z_4}\nn
&&+\gamma^\times_{x_1,z_2,z_1,y_2}G_{-;z_2,z_3}\gamma^\times_{z_3,y_1,x_2,z_4}
+\gamma^{\times}_{z_2,y_1,z_1,y_2}G_{-;z_3,z_2}
\gamma^\times_{x_1,z_3,x_2,z_4})G_{+;z_4,z_5}[-\partial_t+\epsilon_+(-i\partial_x)]_{z_5,z_6}\nn
&&+G_{-;z_6,z_1}(\gamma^\times_{z_1,y_1,x_2,z_2}G_{+;z_2,z_3}
\gamma^\times_{x_1,z_4,z_3,y_2}
+\gamma^{\times}_{z_1,y_1,z_2,y_2}G_{+;z_3,z_2}
\gamma^\times_{x_1,z_4,x_2,z_3}\nn
&&-4\tilde\gamma^\sp_{z_1,z_2,x_1,y_1}G_{-;z_2,z_3}\gamma^\times_{z_3,z_4,x_2,y_2}
-4\gamma^\times_{z_1,z_2,x_2,y_2}G_{-;z_2,z_3}\tilde\gamma^\sp_{z_3,z_4,x_1,y_1})
G_{-;z_4,z_5}[-\partial_t+\epsilon_-(-i\partial_x)]_{z_5,z_6}\}\nonumber\nn
\lambda^2\partial_\lambda\gamma^\sp_{x_1,y_1,x_2,y_2}
&=&-\int_{z_1,\ldots,z_6}\{G_{+;z_6,z_1}(-\gamma^\times_{x_1,y_1,z_1,z_2}G_{+;z_2,z_3}
\gamma^\times_{x_2,y_2,z_3,z_4}
-\gamma^\times_{x_2,y_2,z_1,z_2}
G_{+;z_2,z_3}\gamma^\times_{x_1,y_1,z_3,z_4}\nn
&&+\gamma^\times_{x_1,y_2,z_1,z_2}G_{+;z_2,z_3}
\gamma^\times_{x_2,y_1,z_3,z_4}
+\gamma^\times_{x_2,y_1,z_1,z_2}G_{+;z_2,z_3}
\gamma^\times_{x_1,y_2,z_3,z_4})G_{+;z_4,z_5}[-\partial_t+\epsilon_+(-i\partial_x)]_{z_5,z_6}\nn
&&+G_{-;z_6,z_1}(-16\gamma^\sp_{z_1,z_2,x_1,y_1}
G_{-;z_2,z_3}\gamma^\sp_{z_3,z_4,x_2,y_2}
-16\gamma^\sp_{z_1,z_2,x_2,y_2}G_{-;z_2,z_3}\gamma^\sp_{z_3,z_4,x_1,y_1}\nn
&&+16\gamma^\sp_{z_1,z_2,x_1,y_2}G_{-;z_2,z_3}\gamma^\sp_{z_3,z_4,x_2,y_1}
+16\gamma^\sp_{z_1,z_2,x_2,y_1}G_{-;z_2,z_3}\gamma^\sp_{z_3,z_4,x_1,y_2}\\
&&-4\gamma^\sp_{z_2,y_1,z_1,y_2}G_{-;z_3,z_2}\gamma^\sp_{x_1,z_4,x_2,z_3}
+4\gamma^\sp_{z_2,y_2,z_1,y_1}G_{-;z_3,z_2}\gamma^\sp_{x_1,z_4,x_2,z_3})G_{z_4,z_5}
[-\partial_t+\epsilon_-(-i\partial_x)]_{z_5,z_6}\}.\nonumber
\eea
The Callan-Symanzik strategy can be used again to simplify
the right hand sides as
\bea
\partial_\lambda\gamma^\times_{x_1,y_1,x_2,y_2}
&=&\partial'_\lambda\int_{z_1,\ldots,z_4}[G_{+;z_4,z_1}(-4\gamma^\times_{x_1,y_1,z_1,z_2}
G_{+;z_2,z_3}\tilde\gamma^\sp_{z_3,z_4,x_2,y_2}
+\gamma^{\times}_{z_2,y_1,z_1,y_2}G_{-;z_3,z_2}\gamma^\times_{x_1,z_3,x_2,z_4})\nn
&&+G_{-;z_4,z_1}(\gamma^\times_{z_1,y_1,x_2,z_2}G_{+;z_2,z_3}\gamma^\times_{x_1,z_4,z_3,y_2}
-4\gamma^\times_{z_1,z_2,x_2,y_2}G_{-;z_2,z_3}\tilde\gamma^\sp_{z_3,z_4,x_1,y_1})]\nn
\partial_\lambda\gamma^\sp_{x_1,y_1,x_2,y_2}&=&\partial'_\lambda\int_{z_1,\ldots,z_4}[
G_{+;z_4,z_1}(-\gamma^\times_{x_2,y_2,z_1,z_2}G_{+;z_2,z_3}\gamma^\times_{x_1,y_1,z_3,z_4}
+\gamma^\times_{x_2,y_1,z_1,z_2}G_{+;z_2,z_3}\gamma^\times_{x_1,y_2,z_3,z_4})\nn
&&+G_{-;z_4,z_1}(-16\gamma^\sp_{z_1,z_2,x_1,y_1}G_{-;z_2,z_3}\gamma^\sp_{z_3,z_4,x_2,y_2}
+16\gamma^\sp_{z_1,z_2,x_1,y_2}G_{-;z_2,z_3}\gamma^\sp_{z_3,z_4,x_2,y_1}\nn
&&-4\gamma^\sp_{z_2,y_1,z_1,y_2}G_{-;z_3,z_2}\gamma^\sp_{x_1,z_4,x_2,z_3})].
\eea
where the one-loop beta functions appear on the right hand side.
These relations assume the form
\bea\label{evquart}
\partial_\lambda\gamma^\times_{p,q,r,s}
&=&\partial'_\lambda\int_{u,v}[G_{+;u}(-4\gamma^\times_{p,q,u,v}
G_{+;v}\gamma^\sp_{v,u,r,s}
+\gamma^{\times}_{v,q,u,s}G_{-;v}\gamma^\times_{p,v,r,u})\nn
&&+G_{-;u}(\gamma^\times_{u,q,r,v}G_{+;v}\gamma^\times_{p,u,v,s}
-4\gamma^\times_{u,v,r,s}G_{-;v}\tilde\gamma^\sp_{v,u,p,q})]\nn
\partial_\lambda\gamma^\sp_{p,q,r,s}&=&\partial'_\lambda\int_{u,v}[
G_{+;u}(-\gamma^\times_{r,s,u,v}G_{+;v}\gamma^\times_{p,q,v,u}
+\gamma^\times_{r,q,u,v}G_{+;v}\gamma^\times_{p,s,v,u})\nn
&&+G_{-;u}(-16\gamma^\sp_{u,v,p,q}G_{-;v}\gamma^\sp_{v,u,r,s}
+16\gamma^\sp_{u,v,p,s}G_{-;v}\gamma^\sp_{v,u,r,q}
-4\gamma^\sp_{v,q,u,s}G_{-;v}\gamma^\sp_{p,u,r,v})].
\eea
in momentum space.

\section{Running coupling constant at the Fermi point}
We first confront our evolution equation with the standard one-loop Wilsonian
renormalization group approach to the Luttinger model \cite{shankar,schultz}
where the effective bare coupling constant is supposed to be energy-momentum
independent and its value is given by the coefficient of the
quartic term in the fermion fields taken at the one-shell Fermi-point.
Furthermore we set $\gamma^\sp=0$ at the initial conditions and at any
subsequent stage of the evolution. The evolution of the self-energy, 
Eq. \eq{quadr} is now
\be
\partial_\lambda\Sigma_{\pm;p}=-\partial'_\lambda\int_q\frac{\hat\gamma^\times(\lambda)e^{i\eta q^0}}
{\frac{1}{\lambda}(iq^0-\epsilon_{\mp;\v{q}})-\Sigma_\mp}.
\ee
Because the self-energy corrections are $\ord{\gamma}$, $\Sigma_\pm$
can safely be ignored in the right hand side
in the asymptotically weakly coupled initial conditions, imposed
at $\lambda_0\approx0$. The right hand side starts with a $p$-independent
value at the beginning and obviously keeps this simple form during the rest
of the evolution,
\be
\partial_\lambda\Sigma_{\pm;p}=-\hat\gamma^\times_0(\lambda)\int_q
\frac{e^{i\eta q^0}}{(iq^0-\epsilon_{\mp;\v{q}})}.
\ee
The self energy remains real and it can be absorbed in the renormalization of the
chemical potential.

Let us continue with the evolution of $\gamma^\times(\lambda)$ which, according
to Eq. \eq{evquart} is governed by the equation
\be\label{evquart2}
\partial_\lambda\hat\gamma^\times(\lambda)=\partial'_\lambda\int_{u,v}
\gamma^{\times2}(\lambda)[G_{+;u}G_{-;-u}+(\lambda)G_{-;u}G_{+;u+2k_F}].
\ee
The more detailed form of the two contributions on the right hand side,
\bea\label{tint}
\int G_{+;u}G_{-;-u}&=&-\int_{-\Lambda}^\Lambda\frac{d\v{u}}{2\pi}
\int_{-\infty}^{+\infty}\frac{du^0}{2\pi}\frac{1}{\pa{iu^0-\v{u}}\pa{iu^0+\v{u}}}\nn
\int G_{-;u}G_{+;u+2k_F}&=&\int_{-\Lambda}^\Lambda\frac{d\v{u}}{2\pi}
\int_{-\infty}^{+\infty}\frac{du^0}{2\pi}\frac{1}{\pa{iu^0+\v{u}}\pa{iu^0-\v{u}}},
\eea
shows the vanishing of the beta function,
\be
\partial_\lambda \gamma_0^\times=0,
\ee
the absence of dressing in agreement in Refs. \cite{shankar,schultz}.

Notice that the crucial step of this argument, the formal cancellation of
the two integrals in Eqs. \eq{tint}, takes place at the Fermi points
only. A more careful study of the evolution equation
\eq{evquart} presented in the next Section reveals that the
incomplete cancellation between the contributions on the right hand side of Eq.
\eq{evquart2} generates strongly coupled dynamics in the vicinity of 
the Fermi points.

\section{$\beta$ functions off the Fermi point}
Let us consider now consider the complete evolution.
One side of the problem is the evolution of the propagator, driven by
the quadratic part, Eq. \eq{quadr}. The other one is the evolution of the
two-body interactions, given by the quartic pieces, Eqs. \eq{evquart}.
This solution of this set of coupled integro-differential equations represents
an involved numerical problem, owing to the fact that the evolution of
momentum dependent functions actually involves infinitely many parameters.
We reduce the complexity in this work in order to arrive at an analytically
treatable problem by suppressing the evolution of the quadratic part
in our ansatz \eq{ans}, i.e. we impose $\Sigma=0$, and leave the building up
of the true interactions for the quartic effective vertices.

The quartic evolution, shown by Eqs. \eq{evquart}, is still a formidable
numerical problem due to the singular structures generated in the
integrand and will be simplified in the following manner.
We assume that the naive ideas about the dominance of the interaction
vertices by momentum independent parts is valid and introduce
the usual running coupling constants, given by the strength of
effective interactions at a certain kinematical point. The only
generalization with respect to the previous Section is that the
running coupling constant can be defined off the Fermi points.
We shall then check the consistency of this picture by looking into the magnitude
and the momentum-dependence of the effective two-body vertices
obtained in this approximation.

The $\beta$-functions for the coupling constants
$\gamma^\times_{\tilde p,\tilde q,\tilde r,\tilde s}=\gamma^\times$ and
$\gamma^\sp_{\tilde p,\tilde q,\tilde r,\tilde s}=\gamma^\sp$
at a given point $(\tilde p,\tilde q,\tilde r,\tilde s)$ are introduced as
\bea\label{betafcs}
\beta^\times&=&\beta^\times_{\tilde p,\tilde q,\tilde r,\tilde s}
=\lambda\partial_\lambda{\gamma^\times}_{\tilde p,\tilde q,\tilde r,\tilde s},\nn
\beta^\sp&=&\beta^\sp_{\tilde p,\tilde q,\tilde r,\tilde s}
=\lambda\partial_\lambda\gamma^\sp_{\tilde p,\tilde q,\tilde r,\tilde s}.
\eea
Let us assume that the region around the point $\tilde p,\tilde q,\tilde r,\tilde s$
dominates the dynamics. Then we can approximate the solution of the
integro-differential equations \eq{evquart} by replacing the vertex function
$\gamma_{p,q,r,s}$ by the momentum-independent
$\gamma=\gamma_{\tilde p,\tilde q,\tilde r,\tilde s}$
in the loop integrals. The approximation is better if the
point $(\tilde p,\tilde q,\tilde r,\tilde s)$ is closer to the Fermi point.

We shall parametrize the energy-momentum as
\be
(-i\omega,\v{p})=(\pm\hat{\v{p}}+\Delta_p,\hat{\v{p}}\pm k_F)
\ee
where the plus and minus sign stand for the right and left moving
particles, respectively. The parameters $\Delta_p$, $\Delta_q$, 
$\Delta_r$, and $\Delta_s$ measure the distance from the mass shell. 
The momentum conservation yields
$\hat{\v{s}}-\hat{\v{r}}=\hat{\v{p}}-\hat{\v{q}}$ and the energy conservation
leads to the condition $2(\hat{\v{p}}-\hat{\v{q}})=\Delta_p-\Delta_q+\Delta_r-\Delta_s$
and $0=\Delta_p-\Delta_q+\Delta_r-\Delta_s$
in equations for $\gamma^\times$ and $\gamma^\sp$, respectively.

The beta functions in Eqs. \eq{evquart} can then be written as
\bea\label{betaeq}
\beta^\times&=&-\lambda^2(\beta_\times^{\times\times}\gamma^{\times2}
+\beta_\times^{\times\sp}\gamma^\times\gamma^\sp),\nn
\beta^\sp&=&-\lambda^2(\beta_\sp^{\times\times}\gamma^{\times2}
+\beta_\sp^{\sp\sp}\gamma^{\sp2}),
\eea
with
\bea\label{betacoeff}
\beta_\times^{\times\times}&=&-2\pi\ln\left|\frac{
(2\hat{\v{q}}-\Delta_q+\Delta_r)(2\hat{\v{r}}-\Delta_q+\Delta_r)
[(2\hat{\v{r}}+\Delta_p+\Delta_r)^2-4\Lambda^2]}
{(2\hat{\v{r}}+\Delta_p+\Delta_r)(2\hat{\v{p}}-\Delta_p-\Delta_r)
[(2\hat{\v{q}}+\Delta_q-\Delta_r)^2-4\Lambda^2]}\right|\nn
\beta_\times^{\times\sp}&=&-16\pi\left(\frac{\hat{\v{p}}-\hat{\v{q}}}
{\Delta_p-\Delta_q+2(\hat{\v{p}}-\hat{\v{q}})}
+\frac{\hat{\v{p}}-\hat{\v{q}}}{\Delta_p-\Delta_q}\right)\nn
&=&-8\pi\left(1+\frac{\Delta_r-\Delta_s}{\Delta_p-\Delta_q}\right)
\left(\frac{1}{2+\frac{\Delta_r-\Delta_s}{\Delta_p-\Delta_q}}+1\right)\nn
\beta_\sp^{\times\times}&=&-4\pi\left(\frac{\hat{\v{q}}-\hat{\v{p}}}
{\Delta_q-\Delta_p+2(\hat{\v{q}}-\hat{\v{p}})}
+\frac{\hat{\v{r}}-\hat{\v{q}}+2k_F}{\Delta_r-\Delta_q+2\hat{\v{q}}-2k_F}\right)\nn
\beta_\sp^{\sp\sp}&=&-4\pi\left(8\frac{\hat{\v{q}}-\hat{\v{p}}}
{\Delta_q-\Delta_p+2(\hat{\v{q}}-\hat{\v{p}})}
+8\frac{\hat{\v{r}}-\hat{\v{q}}+2k_F}{\Delta_r-\Delta_q+2\hat{\v{q}}-2k_F}-\frac{\hat{\v{p}}+\hat{\v{r}}+2k_F}
{\Delta_p+\Delta_r+2\hat{\v{r}}+2k_F}\right).
\eea
See Appendix \ref{loopint} for details.

The beta functions display singularities arbitrarily close to the Fermi points 
which lead to singularities in the effective coupling constants \cite{sinec}. 
We shall  integrate out the beta functions in order to see the singularity 
structure in a bit more detailed manner. Two kinematical regions will be considered,
(a): $\hat{\v{q}}=\hat{\v{p}}$, $\Delta=0$, 
$\hat{\v{r}},\hat{\v{p}}\ll k_F,\Lambda$, and 
(b): $\hat{\v{q}}=\hat{\v{p}}$, $2\hat{\v{r}}+\Delta_p+\Delta_r=2\hat{\v{p}}+\Delta_q-\Delta_r$,
$\hat{\v{r}},\hat{\v{p}}\ll k_F$, where
$\beta_\times^{\times\sp}=0$, $\beta_\sp^{\times\times}\approx4\pi$,
$\beta_\sp^{\sp\sp}\approx36\pi$ by ignoring 
$\ord{\hat{\v{p}}/k_F}$, $\ord{\hat{\v{r}}/k_F}$ direction dependent terms and
\be
\beta_\times^{\times\times}\approx\begin{cases}
2\pi\frac{\hat{\v{r}}^2-\hat{\v{p}}^2}{\Lambda^2}&(a),\cr
2\pi\ln\left|\frac{2\hat{\v{p}}+\Delta_q-\Delta_r}{2\hat{\v{p}}-\Delta_q+\Delta_r}\right|
&(b).\end{cases}
\ee
The evolution of $\gamma^{\times}$ is autonomous,
\be
\lambda\frac{d\gamma^{\times}}{d\lambda}=-\lambda^2
\beta_\times^{\times\times}\gamma^{\times2},
\ee
with the solution,
\be\label{indrun}
\gamma^\times(\lambda)=\frac{\gamma^\times_0}
{1+\gamma^\times_0\beta_\times^{\times\times}\lambda^2},
\ee
where $\gamma^\times(0)=\gamma^\times_0$ denotes the bare coupling constant
which is supposed to be small, $|\gamma_0^\times|\ll1$. The solution \eq{indrun}
shows that our truncation resums the geometrical series of the diagonal
part of the vertex function. The truncation of 
the effective action is consistent and the system remains
weakly coupled for $\gamma_0^\times\beta_\times^{\times\times}>0$, in which
case $|\gamma^\times(\lambda)|$ decreases during the evolution. On the contrary, 
for $\gamma_0^\times\beta_\times^{\times\times}<-1$ a Landau-pole is predicted at
\be
\lambda_L=\sqrt{-\frac{1}{\gamma_0\beta_\times^{\times\times}}}<1,
\ee
before the desired bare coupling strength $\gamma_0$ is reached. 
The Landau-pole appears in region (a) due to the cutoff effects when the
inequality
\be
\frac{1}{2\pi}<\gamma_0^\times\frac{\hat{\v{p}}^2-\hat{\v{r}}^2}{\Lambda^2}
\ee
is satisfied by the bare theory. This pole would be absent in a typical
high energy application when the cutoff is large but may appear as a 
non-universal effect when the model is considered as an 
effective theory with low cutoff, $\Lambda\ll k_F$. The Landau-pole
is always present in region (b), in the vicinity of the mass shell
and the Fermi point, where the inequality 
\be
\frac{1}{2\pi}<\gamma_0^\times\ln\left|\frac{1-\frac{\Delta_q-\Delta_r}{2\hat{\v{p}}}}
{1+\frac{\Delta_q-\Delta_r}{2\hat{\v{p}}}}\right|
\ee
can  always be satisfied for an arbitrary value of $\gamma_0^\times$.

The effective coupling constants $\gamma^\sp$ is obtained by integrating the
beta function
\be
\beta^\sp=-4\lambda^2\pi(\gamma^{\times2}+9\gamma^{\sp2}).
\ee
The evolution is autonomous for $\gamma^\times(\lambda)\approx0$ and gives
\be
\gamma^\sp(\lambda)=\frac{\gamma^\sp_0}{1+36\pi\gamma^\sp_0\lambda^2}
\ee
with another Landau-pole when
\be
\gamma^\sp_0<-\frac{1}{36\pi}.
\ee
When $\gamma^\times(\lambda)$ approaches a Landau-pole then
$\gamma^\sp\to-\infty$ because the singularity is non-integrable.

It is worthwhile noting that $\beta_\times^{\times\times}$ and together with it 
the coupling strengths $\gamma^\times(\lambda)$ and $\gamma^\sp(\lambda)$
are scale invariant in regime (b), i.e. the external scale parameters
$k_F$ and $\Lambda$ decouple from the dynamics. As a result, the beta functions,
obtained by means of the usual renormalization group schemes where
the energy-momentum components of subtraction point scale with $\lambda$
in a homogeneous manner, are vanishing. Furthermore, the beta 
functions \eq{betacoeff} and the effective coupling strengths become scale 
invariant in the limit $\Lambda\to\infty$ at the  subtraction point, 
$\omega_p=\omega_q=\omega_r=\omega_s=0$,
$\hat{\v{p}}/3=\hat{\v{q}}=-\hat{\v{r}}=\hat{\v{2}}$, used in Ref. \cite{mdceb}.

\section{Summary}
The transmutation of quasiparticles between the short and the
long distance scales is a challenging problem from particle to
condensed matter physics. We considered one dimensional
massless fermions, the Luttinger model in this work and applied the 
functional form of the internal space renormalization group with
sharp momentum cutoff which yields a functional differential equation for the
effective action. The initial conditions for the differential equation
are known. This equation was truncated in such a manner that the
weakly coupled nature of the interactions can be checked. The solution shows
strong interactions and singular momentum dependence in the vicinity of
the Fermi points in the two-body channels and Landau poles
for arbitrary bare parameters. But our results are rather preliminary,
one should systematically enlarge the ansatz for the effective action
and study the stability of Landau poles in the result.

We argued that cutoff effects, in particular Landau poles, are generated 
by the $\ord{\Lambda^{-2}}$ part of the renormalized action in the normal 
ordered schemes, such as bosonization. We believe that more work is needed
on the one hand, to increase the reliability of the results obtained by
completely regulated bare theories and on the other hand, to clarify
the role the $\ord{\Lambda^{-2}}$ interactions may play in the bosonized
theory.

\begin{acknowledgments}
We thank Walter Metzner for his useful remarks about this work.
\end{acknowledgments}

\appendix
\section{Notation, conventions}\label{not}
The space-time and Fourier-space integrals are defined as
\be
\int_x=\int d^2x=\int d\v{x}dt=a_sa_t\sum_x,~~~~
\int_p=\int\frac{d^2p}{(2\pi)^2}=\sum_p,~~~~
f_x=\int_pf_pe^{ipx},
\ee
where $x=(t,\v{x})$, $p=(\omega,\v{p})$. In Euclidean space-time
one has $px=p^0x^0+\v{p}\v{x}$. The letters $a$, $b,\ldots$ and
$p$, $q,\ldots$ denote space-time or energy-momentum variables, respectively.
The integration is sometime shown as scalar product, $f\cdot g=\int_xf_xg_x$.

The functional derivative is defined in $d$-dimensional space-time lattice as
\be
\fd{}{\Psi_x}=\frac{1}{a_sa_t}\frac{\partial}{\partial\Psi^L_x}
\ee
where $\Psi^L$ is the lattice field variable whose dimension is removed by
the lattice spacing and the factor $1/a_sa_t$ is needed in order to satisfy
the relation
\be
\fd{}{\Psi_y}\int_xf(\phi_x)=f'(\Psi_y).
\ee
In an analogous manner we have
\be
\fd{}{\Psi_p}=V\frac{\partial}{\partial\Psi^L_p}.
\ee

\section{One-loop integrals}\label{loopint}
This Appendix contains some details concerning the computation of the
one-loop integrals of Eqs. \eq{evquart} by means of sharp momentum space
cutoff. In order to simplify the  results we assume that the absolute 
value of the momenta, counted from the Fermi point does not exceed $\Lambda/4$.

\subsection{$\gamma^\times$}
For the first equation we need the following integrals,
\bea
I_1&=&\frac{1}{\lambda^2}\int_{\v{t},\omega} G_{+;t}G_{+;p-q+t},\nn
I_2&=&\frac{1}{\lambda^2}\int_{\v{t},\omega} G_{+;t}G_{-;p+r-t},\nn
I_3&=&\frac{1}{\lambda^2}\int_{\v{t},\omega} G_{-;t}G_{+;t-q+r},\nn
I_4&=&\frac{1}{\lambda^2}\int_{\v{t},\omega} G_{-;t}G_{-;t+r-s},
\eea
where
\be
G_\sigma=\frac{\lambda}{-\partial_t+\epsilon_\sigma(-i\partial_x)}.
\ee
One uses the residuum theorem in integrating over the frequency in the integral
\be
I_1=\int_{-\infty}^\infty d\v{u}\int_{-\infty}^{+\infty}d\omega
\frac{\rho_\Lambda(\v{u})\rho_\Lambda(\v{u}+\v{p}-\v{q})}
{\pa{\omega+i\v{u}}\pa{\omega+\omega_p-\omega_q+i(\v{u}+\v{p}-\v{q})}}
\ee
and finds
\be
I_1=2i\pi\int_{-\infty}^\infty d\v{u}\frac{\rho_\Lambda(\v{u})\rho_\Lambda(\v{u}+\v{p}-\v{q})
[\Theta(-\v{u})-\Theta(-\v{u}+\v{q}-\v{p})]}
{\omega_p-\omega_q+i(\v{p}-\v{q})}
\ee
which results
\be
I_1=\frac{2\pi(\hat{\v{p}}-\hat{\v{q}})}{\Delta_p-\Delta_q+2(\hat{\v{p}}-\hat{\v{q}})}.
\ee

The other three integrals can be computed in a similar manner. For
\be
I_2=\int_{-\infty}^\infty d\v{u}\int_{-\infty}^{+\infty}d\v{\omega}
\frac{\rho_\Lambda(\v{u})\rho_\Lambda(\v{u}-\v{p}-\v{r})}
{\pa{\omega+i\v{u}}\pa{\omega-\omega_p-\omega_r-i(\v{u}-\v{p}-\v{r})}}
\ee
we find
\be
I_2=F_2(0)-F_2(-\Lambda)-[F_2(\Lambda)-F_2(\hat{\v{p}}+\hat{\v{r}})]
\ee
for $|\hat{\v{p}}|,|\hat{\v{r}}|<\Lambda/4$ where
\be
F_2(\v{u})=-\pi\ln\left|\v{u}-\frac{\hat{\v{p}+\hat{\v{r}}}}{2}
-i\frac{\omega_p+\omega_r}{2}\right|
=-\pi\ln\left|\v{u}-\hat{\v{r}}-\frac{\Delta_p+\Delta_r}{2}\right|.
\ee
The final result is
\be
I_2=-\pi\ln{\left|\frac{\pa{\hat{\v{r}}+\frac{\Delta_p+\Delta_r}{2}}
\pa{ \hat{\v{p}}-\frac{\Delta_p+\Delta_r}{2}}}
{\pa{\Lambda-\hat{\v{r}}-\frac{\Delta_p+\Delta_r}{2}}
\pa{\Lambda+ \hat{\v{r}}+\frac{\Delta_p+\Delta_r}{2}}}\right|}.
\ee

The integral
\be
I_3=-\int_{-\infty}^\infty d\v{u}\int_{-\infty}^{+\infty}d\v{\omega}
\frac{\rho_\Lambda(\v{u})\rho_\Lambda(\v{u}-\v{q}+\v{r})}
{\pa{\omega-i\v{u}}\pa{\omega-\omega_q+\omega_r+i(\v{u}-\v{q}+\v{r}-2k_F)}}
\ee
becomes
\be
I_3=-F_3(\Lambda)-F_3(-\Lambda)+F_3(0)+F_3(\hat{\v{q}}-\hat{\v{r}})
\ee
where
\be
F_3(\v{u})=\pi\ln\left|\v{u}+\frac{\hat{\v{r}}-\hat{\v{q}}}{2}
+i\frac{\omega_q-\omega_r}{2}\right|
\ee
after the integration over the frequency. Trivial steps give
\be
I_3=\pi\ln{\left|\frac{\pa{ \frac{\Delta_q-\Delta_r}{2}- \hat{\v{q}}}
\pa{ \frac{\Delta_q-\Delta_r}{2}-\hat{\v{r}}}}
{\pa{\hat{\v{q}}+\frac{\Delta_q-\Delta_r}{2}}^2-\Lambda^2}\right|}.
\ee
Finally, the last integral,
\be
I_4=\int_{-\infty}^\infty d\v{u}\int_{-\infty}^{+\infty}d\v{\omega}
\frac{\rho_\Lambda(\v{u})\rho_\Lambda(\v{u}-\v{s}+\v{r})}
{\pa{\omega-i\v{u}}\pa{\omega+\omega_r-\omega_s+i(-\v{u}-\v{r}+\v{s})}}
\ee
gives
\bea
I_4&=&2i\pi\int_{-\Lambda}^{+\Lambda}d\v{u}\left[
\frac{\Theta(\v{u})}{\omega_r-\omega_s+i(\v{s}-\v{r})}
+\frac{\Theta(\v{u}+\v{r}-\v{s})}{\omega_s-\omega_r-i(\v{s}-\v{r})}\right]\nn
&=&\frac{2\pi(\hat{\v{s}}-\hat{\v{r}})}{\Delta_s-\Delta_r+2(\hat{\v{s}}-\hat{\v{r}})}.
\eea

\subsection{$\gamma^\sp$}
For the second equation in Eqs. \eq{evquart} we need the following five integrals,
\bea
J_1&=&\frac{1}{\lambda^2}\int_{\v{t},\omega} G_{+;t}G_{+;r-s+t},\nn
J_2&=&\frac{1}{\lambda^2}\int_{\v{t},\omega} G_{+;t}G_{+;r-q+t},\nn
J_3&=&\frac{1}{\lambda^2}\int_{\v{t},\omega} G_{-;t}G_{-;p+r-t},\nn
J_4&=&\frac{1}{\lambda^2}\int_{\v{t},\omega} G_{-;t}G_{-;t-r+s},\nn
J_5&=&\frac{1}{\lambda^2}\int_{\v{t},\omega} G_{-;t}G_{-;u-r+q}.
\eea
Due to the left and right symmetry properties $J_4=J_1$ and $J_5=J_2$.
The same procedure than in the case of the previous integrals gives
\bea
J_1&=&\int_{-\infty}^\infty d\v{u}\int_{-\infty}^{+\infty}d\v{\omega}
\frac{\rho_\Lambda(\v{u})\rho_\Lambda(\v{u}-\v{s}+\v{r})}
{\pa{\omega+i\v{u}}\pa{\omega+\omega_r-\omega_s+i(\v{u}+\v{r}-\v{s})}}\nn
&=&2i\pi\int_{-\infty}^\infty d\v{u}\rho_\Lambda(\v{u})\rho_\Lambda(\v{u}-\v{s}+\v{r})
\left[\frac{\Theta(-\v{u})}{\omega_r-\omega_s+i(\v{r}-\v{s})}
+\frac{\Theta(-\v{u}+\v{s}-\v{r})}{\omega_s-\omega_r-i(\v{r}-\v{s})}\right]\nn
&=&\frac{2\pi(\hat{\v{r}}-\hat{\v{s}})}{\Delta_r-\Delta_s +2(\hat{\v{r}}-\hat{\v{s}})},
\eea
and
\bea
J_2&=&-\int_{-\infty}^\infty d\v{u}\int_{-\infty}^{+\infty}d\v{\omega}
\frac{\rho_\Lambda(\v{u})\rho_\Lambda(\v{u}-\v{q}+\v{r})}
{\pa{\omega+i\v{u}}\pa{\omega+\omega_r-\omega_q+i(\v{u}+\v{r}-\v{q})}}\nn
&=&-2i\pi\int_{-\infty}^\infty d\v{u}\rho_\Lambda(\v{u})\rho_\Lambda(\v{u}-\v{s}+\v{r})
\left[\frac{\Theta(-\v{u})}{\omega_r-\omega_q+i(\v{r}-\v{q})}
+\frac{\Theta(-\v{u}-\v{r}+\v{q})}{\omega_q-\omega_r-i(\v{r}-\v{q})}\right]\nn
&=&2\pi\frac{\hat{\v{r}}-\hat{\v{q}}+2k_F}{\Delta_r-\Delta_q+2\hat{\v{q}}-2k_F}
\eea
Finally, the last independent integral is
\bea
J_3&=&-\int_{-\infty}^\infty d\v{u}\int_{-\infty}^{+\infty}d\v{\omega}
\frac{\rho_\Lambda(\v{u})\rho_\Lambda(\v{u}-\v{p}-\v{r})}
{\pa{\omega-i\v{u}}\pa{\omega-\omega_p-\omega_r-i(\v{u}-\v{p}-\v{r}-2k_F)}}\nn
& =&-2i\pi\int_{-\infty}^\infty d\v{u}\rho_\Lambda(\v{u})\rho_\Lambda(\v{u}-\v{p}-\v{r})
\left[\frac{\Theta(\v{u})}{-\omega_p-\omega_r-i(-\v{p}-\v{r}-2k_F)}
+\frac{\Theta(\v{u}-\v{p}-\v{r}-2k_F)}{\omega_p+\omega_r+i(-\v{p}-\v{r}-2k_F)}\right]\nn
&=&-2\pi\frac{\hat{\v{p}}+\hat{\v{r}}+2k_F}{\Delta_p+\Delta_r+2\hat{\v{r}}+2k_F}
\eea

\end{document}